\documentstyle[preprint, aps]{revtex}

\newcommand{\Lag}{{\cal L}}
\newcommand{\DD}{_{{\scriptscriptstyle{1/2}}}}
\newcommand{\LagT}{{\cal L}_{{\scriptscriptstyle{\rm TR}}}}
\newcommand{\LagQ}{{\cal L}_{{\scriptscriptstyle{\rm QED}}}}
\newcommand{\STR}{S_{{\scriptscriptstyle{\rm TR}}}}

\newcommand{\ab}[1]{\bar{#1}}
\newcommand{\Four}[1]{\widetilde{#1}}
\newcommand{\prop}{\Delta_F}
\newcommand{\propF}{\Four{\prop}}
\newcommand{\ldp}{\vbox{\ialign{\hfil$##$\hfil\crcr
 \scriptscriptstyle\leftrightarrow\crcr\noalign{\kern.5pt\nointerlineskip}
 \partial\crcr}}}

\begin{document}

\title{Regularization of QED by a generalized 't Hooft and Veltman method}

\author{Marijan Ribari\v c and Luka \v Su\v ster\v si\v c\footnote{Corresponding author. Phone +386 61 177 3258; fax +386 61 123 1569; electronic address: \tt luka.sustersic@ijs.si\rm.} }

\address{Jo\v zef Stefan Institute, p.p.3000, 1001 Ljubljana, Slovenia}

\maketitle

\begin{abstract}Generalizing the 't Hooft and Veltman method of unitary regulators, we demonstrate for the first time the existence of local, Lorentz-invariant, physically motivated Lagrangians of quantum-electrodynamic phenomena such that: (i)~Feynman diagrams are finite and equal the diagrams of QED but with regularized propagators. (ii)~N-point Green functions are causal. (iii)~$S$-matrix relates only electrons, positrons and photons, is unitary and Lorentz-invariant, and conserves charge and total four-momentum. 
\end{abstract}
\vskip.3in

\section{Introduction}

Perturbative predictions about quantum-electrodynamic phenomena implied by a QED Lagrangian can be computed using the Feynman rules, a regularization method to circumvent ultraviolet divergencies, and a renormalization scheme. Regularization method results in regularized n-point Green functions; a suitable limiting procedure (a renormalization scheme) then leads to physically sensible predictions that are independent of the particular regularization method used. But no known regularized n-point Green functions can be regarded as being based on a physically plausible theory of quantum-electrodynamic phenomena since the derivation of each disregards some of the basic tenets of conventional physics (e.g., by lacking a Lagrangian, by not being Lorentz-invariant, by introducing particles with wrong metric or statistics\ldots). \it So the perturbative predictions of QED presently cannot be derived from a physically plausible theory\rm; for a history of, and comments on this basic problem see, e.g.,\cite{Cao}. Dirac\cite{Dirac} believed that removal of this conceptual inconsistency may lead to an important advance in field theories. 

To show that one can remove this inconsistency \it already in four-dimensional space-time, \rm we will introduce a new, physically motivated modification of the QED Lagrangian and consider it within the theoretical framework of 't Hooft and Veltman\cite{Hooft}: They avoid canonical formalism and take diagrams as the basis from which everything must be derived; so they give a perturbative definition of the S-matrix directly in terms of diagrams corresponding to a given Lagrangian as specified by \it postulated \rm Feynman rules. The question is: How do we modify the QED Lagrangian so that the resulting regularized S-matrix is derived from a theory that is physically plausible?

Gupta\cite{Gupta} has shown already in 1952 that one can modify the QED Lagrangian so that the new Lagrangian results in the S-matrix of QED regularized by certain Pauli-Villars method. And twenty years later 't Hooft and Veltman\cite{Hooft} introduced the method of unitary regulators (HV-method) that (i)~is a variant of Pauli-Villars methods for regularizing propagators, (ii)~requires only an exceedingly simple modification of the initial Lagrangian, and (iii)~is very suitable for proving the causality of the regularized $n$-point Green functions and the unitarity of the resulting S-matrix. Unfortunately both methods introduce also unphysical particles with wrong metric or statistics. To get rid of this serious conceptual deficiency, we will generalize the HV-method.

We will demonstrate the utility of the generalized HV-method by using it to show that there is a perturbative S-matrix of quantum-electrodynamic phenomena derived from a theory that is \it physically plausible \rm(a pp-theory, for short) \it in the following sense: \rm
\begin{itemize}
\item[(i)]A pp-theory of quantum-electrodynamic phenomena is specified in a continuous, four-dimensional space-time by a local, Lorentz-invariant, physically motivated modification of a QED Lagrangian. 
\item[(ii)]The Feynman rules for this modified Lagrangian, \it defined \rm as specified by 't Hooft and Veltman\cite{Hooft}, result in Feynman diagrams that are finite and equal the diagrams of QED but with regularized propagators. 
\item[(iii)]All constants of a pp-theory are measurable in principle; \it there are no auxiliary parameters. \rm 
\item[(iv)]For certain values of these constants, the QED propagators are the same kind of low-energy approximations to their regularizations as required for renormalization by Bogoliubov and Shirkov\cite{Bogol} in certain variants of the Pauli-Villars method. 
\item[(v)]The n-point Green functions of a pp-theory, \it defined \rm as specified by 't Hooft and Veltman\cite{Hooft} in terms of Feynman diagrams, are causal. 
\item[(vi)]The perturbative S-matrix of a pp-theory, \it defined \rm as specified by 't Hooft and Veltman\cite{Hooft} in terms of n-point Green functions, is unitary, Lorentz-invariant, charge and total four-momentum conserving, and relates \it only physical electrons, positrons and photons. \rm
\end{itemize}
Such a pp-theory of quantum-electrodynamic phenomena is not yet known; we cannot incorporate a finite-cutoff, Pauli-Villars, dimensional, or lattice regularization of QED in a pp-theory.

\section{Lorentz-invariant regulators without singularities}

As in the HV-method to each additional singularity of a regularized Feynman propagator corresponds an additional particle\cite{Hooft}, we will first show how we can Lorentz-invariantly regularize Feynman propagators so that they do not acquire additional singularities and have the K\"all\'en-Lehman representation used in proving causality and unitarity\cite{Hooft,Veltm}. Regarding metric and other conventions we follow Refs.~\onlinecite{Hooft,Veltm}; in particular, a four-vector $k = (\vec{k}, ik^0)$, and $k^2 \equiv \vec{k}\cdot \vec{k} - (k^0)^2$.

Consider a Lorentz-invariantly regularized spin 0 Feynman propagator, say, $\prop(x)$ whose space-time Fourier transform
\begin{equation}
   (2\pi)^4 i\propF(k) = (k^2 + m^2 - i\epsilon)^{-1} 
      + \varphi(k^2 - i\epsilon)
   \label{scpro}
\end{equation}
where: (a)~the regulator $\varphi(z)$ is an analytic function of complex variable $z$ with a finite discontinuity somewhere across the negative real axis; (b)~$|(z+m^2)^{-1} + \varphi(z)| < A |z|^{-r}$ for $|z| > 2m^2$, with $r > 1$; (c)~$\varphi(z)$ is real if $\Re z > 0$ and $\Im z = 0$; (d)~$\varphi(z)$ depends on some constant $\Lambda$ so that for any $\Lambda \ge \Lambda_0 > 0$ it has properties (a) to (c) with $A$ independent of $\Lambda$, and, as $\Lambda \to \infty$,
\begin{equation}
   \sup_{\Re z \ge 0, \Im z = 0} |\varphi(z)| \to 0 
   \quad\hbox{and}\quad
   \sup_{|z| < z_0} |\varphi(z)| \to 0
   \qquad\hbox{for any}\quad z_0 > 0 \,.
   \label{regpr}
\end{equation}
As a consequence, the spin 0 propagator provides such a low-energy approximation to its regularization (\ref{scpro}) as required for renormalization by Bogoliubov and Shirkov\cite{Bogol} in certain variants of the Pauli-Villars method.

Using Cauchy's integral formula we can conclude that the Lorentz-invariant regularization (\ref{scpro}) of the spin 0 Feynman propagator admits the K\"all\'en-Lehman representation
\begin{equation}
   (2\pi)^4 i \propF(k) = \int_0^\infty {\rho(s)\over k^2 
       + s - i\epsilon}\, ds 
   \label{KLrep}
\end{equation}
with
\begin{equation}
   \rho(s) = \delta(s-m^2) + (2\pi i)^{-1} \lim_{y \to 0} [
      \varphi(-s-iy) - \varphi(-s+iy) ] \,,
   \label{KLrho}
\end{equation}
$s, y > 0$. Note that $\rho(s)$ is real, $\rho(s) = O(s^{-r})$ as $s \to \infty$, and
\begin{equation}
   \int_0^\infty s^m \rho(s)\, ds = 0
   \qquad\hbox{for}\quad m = 0, 1,\ldots < r-1 \,.
   \label{KLcon}
\end{equation}
So we can decompose the regularized spin 0 propagator $\prop(x)$ into positive and negative energy parts: $\prop(x) = \Theta(x_0) \Delta^+(x) + \Theta(-x_0) \Delta^-(x)$\cite{Hooft}.

The function $-i (2\pi)^{-4} ( \sqrt{\Lambda^2 - m^2} + \Lambda)^n (k^2 + m^2 - i\epsilon)^{-1} (\sqrt{k^2 + \Lambda^2 - i\epsilon} + \Lambda )^{-n}$, $\Lambda > m$, $n = 1, 2, \ldots $, is an example of a Lorentz-invariantly regularized spin 0 Feynman propagator that satisfies the above conditions with $r = n/2 + 1$. Unfortunately, we cannot use such propagators for a physically plausible regularization of QED since we do not know how to construct the corresponding \it local, \rm Lorentz-invariant Lagrangians.

A regulator that satisfies conditions (a)-(c) is by (\ref{KLrep}) a generalization of the Pauli-Villars regulator that has a continuous mass spectrum. Thus, to use such regulators to construct a pp-theory, we have to extend the 't Hooft-Veltman construction of Lagrangians in HV-method\cite{Hooft} to an \it infinite number \rm of additional fields. To provide an example of how this can be done, we will present a local, Lorentz-invariant Lagrangian whose propagators for \it interacting \rm fields can be taken as spin 1 and spin $1\over 2$ propagators regularized so that they acquire no additional singularities and have the K\"all\'en-Lehman representation.

\section{Lagrangian that regularizes QED propagators}

Following Veltman \cite{Veltm}, we will consider QED with massive photons in unitary gauge. Its Lagrangian reads
\begin{equation}
   \LagQ = -{\textstyle{1\over 4}} (\partial_\mu A_\nu - \partial_\nu A_\mu)^2 - {\textstyle{1\over 2}} \mu^2 A^2 -
	\ab{\psi} (\gamma^\mu \ldp_\mu + m )\psi + ie \ab{\psi} \gamma^\mu\psi A_\mu + A_\mu J_\mu + \ab{J_e}\psi 
        + \ab{\psi}J_e\,,
   \label{QEDL}
\end{equation}
where $J_\mu(x)$, $\ab{J_e}(x)$, and $J_e(x)$ are four-vector and bispinor source fields, and the non-vanishing photon mass $\mu < 2 \times 10^{-16}$ eV \cite{Parti}. The Feynman propagators for the four-vector field $A_\mu(x)$ and for the bispinor field $\psi(x)$ are: 
\begin{equation}
   -i (2\pi)^{-4} { \delta_{\mu\nu} + \mu^{-2} k_\mu k_\nu
	\over k^2 + \mu^2 - i\epsilon } \,,\qquad
   -i (2\pi)^{-4} { -i \gamma^\mu k_\mu + m \over k^2 + m^2 
	- i\epsilon } \,.
   \label{PDeq}
\end{equation}
We could use $\LagQ$ to define a pp-theory as specified in Section~I, were the propagators (\ref{PDeq}) faster decreasing when $k^2$ tends to infinity.

However, one can modify the QED Lagrangian (\ref{QEDL}) so that the propagators for the fields $A_\mu$ and $\psi$ are such regularizations of propagators (\ref{PDeq}) that have no additional singularities. Take, for \it example, \rm the following real-valued, local, Lorentz-invariant Lagrangian:
\begin{mathletters}
\label{TTL}
\begin{equation}
   \LagT = - \Lag_1 - \Lag\DD + ie\ab{\psi} \gamma^\mu \psi A_\mu 
         + A_\mu J_\mu + \ab{J_e}\psi + \ab{\psi}J_e
   \label{TTLdf}
\end{equation}
with
\begin{eqnarray}
   \Lag_1 &\equiv& q_1^{-1} \int d^4p\, \Psi_\mu'(x,-p)
      [ \Lambda t(p^2) + p^\nu \ldp_\nu ] \Psi^\mu(x,p)
      \nonumber\\
   &&\qquad{}+ q_1^{-1} s_1 \int d^4p\, d^4p'\, f({p'}^2) f(p^2)
      [ \Psi_\mu'(x, -p')\Psi^{\prime\mu}(x,p) + p'_\nu p^\nu \Psi_\mu
        (x,-p') \Psi^\mu(x,p)
      \nonumber\\
   &&\qquad\quad{}- p^\mu \Psi_\mu(x,-p') {p'}^\nu \Psi_\nu(x,p) ] \,,
      \label{vecL}\\
   \Lag\DD &\equiv& q\DD^{-1} \int d^4p\, \ab{\Psi}\DD(x,-p)
      [ \Lambda t(p^2) + p^\mu\ldp_\mu ] \Psi\DD(x,p)
      \nonumber\\
      &&\qquad{}- q\DD^{-1} s\DD \int d^4p'\, d^4p\, f({p'}^2)
        f(p^2) [ \ab{\Psi}\DD(x,-p') \gamma^\mu \Psi\DD(x,p) p_\mu
         + {\rm c.c.} ] \,,
   \label{spinL}
\end{eqnarray}
\begin{equation}
   A_\mu(x) \equiv \int d^4p f(p^2) \Psi_\mu(x,p) \,,\qquad
   \psi(x) \equiv \int d^4p f(p^2) \Psi\DD(x,p) \,,
   \label{macvar}
\end{equation}
\end{mathletters}
where: $\Psi_\mu(x,p)$ and $\Psi_\mu'(x,p)$ are four-vector-valued functions of two four-vectors $x$ and $p$; $\Psi\DD(x,p)$ is a bispinor-valued function of $x$ and $p$; $2 a \ldp_\mu b \equiv a (\partial_\mu b ) - (\partial_\mu a) b $; $\ab{\Psi}\DD \equiv \Psi\DD^\dagger \gamma^4$; $t(p^2)$ and $f(p^2)$ are real-valued functions of real $p^2$, $\int d^4 p f^2(p^2) = 1$; $q_1$, $s_1$, $q\DD$, $s\DD$, and $\Lambda$ are real constants.

There are three kinds of reasons for the chosen form (\ref{TTL}) of the Lagrangian $\LagT$:
\begin{itemize}
\item[(A)]It is our purpose to show that there are Lagrangians that generalize the t'Hooft and Veltman method of unitary regulators \cite{Hooft} to an infinite number of additional fields \it but do not introduce additional particles. \rm So we constructed the Lagrangian $\LagT$ modifying $\LagQ$ on the analogy of HV-method\cite{Hooft}: (i)~We introduced an infinite number of four-vector and bispinor fields of $x$ that have a continuous index $p$, namely $\Psi_\mu(x,p)$, $\Psi'_\mu(x,p)$, and $\Psi\DD(x,p)$. (ii)~We replaced the free part of $\LagQ$ with the free Lagrangian of these fields,  $-\Lag_1 - \Lag\DD$, which is of the first order in $\partial$ and has a nondiagonal mass matrix. (iii)~In the interaction and source terms of $\LagQ$, we replaced the fields $A_\mu(x)$ and $\psi(x)$ with weighted integrals (\ref{macvar}) of $\Psi_\mu(x,p)$ and $\Psi\DD(x,p)$ over the continuous index $p$. 
\item[(B)]We tried to simplify the calculations of regularized propagators. We could do without the four-vector function $\Psi'_\mu(x,p)$ which we introduced solely to be able to use the same functions $t(p^2)$ and $f(p^2)$ in $\Lag_1$ and $\Lag\DD$. We introduced $\ldp$ to make $\LagT$ itself real-valued, not only its action real as required.
\item[(C)]The Euler-Lagrange equations of $\LagT$ resemble the Boltzmann integro-differential transport equation, which can better model rapidly varying, ``ultra-high-energy'', macroscopic fluid phenomena than the differential equations of motion of fluid dynamics. (To this end it uses an infinite number of fields to take some  account of the underlying microscopic behaviour.) Euler-Lagrange equations of $\LagT$ may be regarded as classical transport equations of motion for the one-particle distribution of some infinitesimal entities, such as X-ons surmised to underly all physical phenomena by Feynman\cite{Feynm}. Which all are physical motivations for the type of Lagrangian we constructed; we discuss them in detail in Ref.~\cite{mi001}.
\end{itemize}

Using the Euler-Lagrange equations of $\LagT$ and proceeding as in Ref.~\onlinecite{mi002}, we calculate the dependence of $\Psi_\mu(x,p)$ and  $\Psi'_\mu(x,p)$ on $J_\mu(x)$, and of $\Psi\DD(x,p)$ on $J_e(x)$. Thereby we can infer that the Feynman propagator for the four-vector field $A_\mu(x)$ defined by (\ref{macvar}) equals
\begin{equation}
   -i (2\pi)^{-4} \Four{g}_1 \, {\delta_{\mu\nu} + \Four{\mu}^{-2} k_\mu k_\nu\over k^2 + \Four{\mu}^2 } \,,
	\label{phpro}
\end{equation}
\begin{equation}
   \Four{g}_1(k^2) \equiv q_1 s_1^{-2} I_{10} I_{20}^{-2}\,, \qquad
   \Four{\mu}(k^2) \equiv |s_1|^{-1} I_{20}^{-1} \,,
\end{equation}
where $I_{mn}(k^2)$ is an analytic function of the complex variable $k^2$ such that
\begin{equation}
   I_{mn}(k^2) = 2 \pi^2 \Lambda^{-m} \int_0^\infty y^{m+n}
      f^2(y) t^{-m}(y) [ \sqrt{ 1 + \Lambda^{-2} k^2 y t^{-2}(y)}
      + 1]^{-m} dy
   \label{Imn} 
\end{equation}
for $k^2 > 0$; and the Feynman propagator for the bispinor field $\psi(x)$ defined by (\ref{macvar}) equals
\begin{equation}
   -i (2\pi)^{-4} \Four{g}\DD \, { - i\gamma^\mu k_\mu + \Four{m} \over
       k^2  + \Four{m}^2 } \,,
   \label{elpro}
\end{equation}
\begin{equation}
   \Four{g}\DD(k^2) \equiv q\DD s\DD^{-1} I_{10}I_{20}^{-1}\,, 
      \qquad \Four{m}(k^2) \equiv s\DD^{-1} \{ 1 - s\DD^2 [
      I_{10} I_{11}  + {\textstyle{1\over4}} k^2 I_{20}^2 ] \}
      I_{20}^{-1} \,;
\end{equation}
where $k^2$ has to be replaced everywhere with $k^2 - i\epsilon$, by the Feynman prescription.

If functions $t(p^2)$ and $f(p^2)$ are such that
\begin{equation}
   \int_0^\infty f^2(y) t(y) \, | \sqrt y /t(y)|^{l+1} dy = 0
   \label{ftpog}
\end{equation}
for $l=0, -1, \ldots,-n$, then for complex values of $k$ as $|k^2| \to \infty$:
\begin{equation}
   \left| \Four{g}_1 {\delta_{\mu\nu} + \Four{\mu}^{-2} k_\mu k_\nu
	\over k^2 + \Four{\mu}^2} \right| = O(|k^2|^{(1-n)/2}) \,,
   \qquad
   \left| \Four{g}\DD \, {\Four{m} - i\gamma^\mu k_\mu 
      \over k^2 + \Four{m}^2 } \right| = O(|k^2|^{-n/2}) \,.
   \label{phreg}
\end{equation}

When the function $y/t^2(y)$ takes only a finite number of real values $v_i$, $i = 1, 2, \ldots$, we can explicitly evalute integrals (\ref{Imn}); we obtain
\begin{equation}
   I_{mn}(k^2) = \Lambda^{-m} \sum_{i} A_{mni} v_i^{m} [ \sqrt{ 1 + \Lambda^{-2} v_i k^2 } + 1]^{-m} \,,
   \label{Imnexample} 
\end{equation}
where $A_{mni}$ are real constants. Considering such a case, we can show that for any $\mu^2$, $m^2$ and integer $n$, there exist functions $f(p^2)$ and $t(p^2)$, and constants $s_1$, $s\DD$, $q_1$, $q\DD$, and $\Lambda_0 > 0$ such that the propagators (\ref{phpro}) and (\ref{elpro}) with $\Lambda > \Lambda_0$ are regularizations of spin 1 and spin $1\over 2$ propagators (\ref{PDeq}) such that: (i)~they have properties analogous to those of propagator (\ref{scpro}), and (ii)~there is a positive constant $k_0^2$ such that for all $k^2 \ge -\Lambda^2 k_0^2$ the functions $I_{mn}(k^2)$, $\Four{g}\DD(k^2)$, $\Four{\mu}(k^2)$, $\Four{g}_1(k^2)$, and $\Four{m}(k^2)$ are real. In such a case: (i)~The constants $s_1$, $s\DD$, $q_1$, and $q\DD$ are such that
\begin{equation}
   \Four{\mu}^2(-\mu^2) = \mu^2 \,, \qquad
   \Four{m}^2(-m^2) = m^2 \,,
   \label{cond1}
\end{equation}
\begin{equation}
   \Four{g}\DD(-\mu^2) = 1 + d\Four{\mu}^2(k^2 = -\mu^2)/ dk^2 \,,
   \qquad
   \Four{g}_1(-m^2) = 1 + d\Four{m}^2(k^2 = -m^2)/ dk^2 \,.
   \nonumber
\end{equation}
So the propagators (\ref{phpro}) and (\ref{elpro}) have poles at $k^2 = -\mu^2 $ and $k^2 = - m^2$, where their behaviour is given by the spin 1 and spin $1\over 2$ propagators (\ref{PDeq}) with $\epsilon = 0$. (ii)~The difference between spin 1 propagator and propagator (\ref{phpro}) depends on the value of $\Lambda$ so that it satisfies relations analogous to (\ref{regpr}); and the same goes for spin $1\over 2$ propagators. (iii)~The propagators (\ref{phpro}) and (\ref{elpro}) are analytic functions of $k^2$ that (a)~are not continuous everywhere across the negative real axis, (b)~have no additional singularities to those of spin 1 and spin $1\over 2$ propagators (\ref{PDeq}), and (c)~satisfy relations (\ref{phreg}). For any integer $n \ge 3$, their K\"all\'en-Lehman integral representations are superconvergent: in $x$-space we can decompose the Feynman propagators (\ref{phpro}) and (\ref{elpro}) into positive and negative energy parts without contact terms\cite{Hooft}.

\section{Physically plausible regularization of QED}

To obtain a perturbative S-matrix of quantum-electrodynamic phenomena based on the Lagrangian $\LagT$, say $\STR$, we use the 't Hooft-Veltman definition of an S-matrix\cite{Hooft}. In view of results of Sec.III, there are functions $f(p^2)$ and $t(p^2)$, and constants $s_1$, $s\DD$, $q_1$, $q\DD$, and $\Lambda$ such that the n-point Green functions of $\LagT$ and the corresponding S-matrix $\STR$ have the following properties: 
\begin{itemize}
\item[(i)]As the Lagrangian $\LagT$ has the same interaction and source terms as the QED Lagrangian $\LagQ$, they are expressed in terms of QED diagrams with the spin 1 and spin $1\over 2$ propagators (\ref{PDeq}) replaced with their regularizations (\ref{phpro}) and (\ref{elpro}), whereas the vertices are the same as in QED, i.e., $(2\pi)^4 \gamma_\mu$; so all diagrams are finite.
\item[(ii)]As the propagators (\ref{phpro}) and (\ref{elpro}), and higher-order 2-point Green functions of $\LagT$ have no additional singularities, $\STR$ relates the same particles as the S-matrix of QED with massive photons in unitary gauge: electrons and positrons, each with two possible polarization vectors, and massive photons with three possible polarization vectors; none of them with wrong metric or statistics. 
\item[(iii)]N-point Green functions are causal and $\STR$ is unitary to any order in the fine structure constant. To show this we may follow the t Hooft-Veltman proof for the HV-method\cite{Hooft}, since the propagators (\ref{phpro}) and (\ref{elpro}) admit the K\"all\'en-Lehman representation. \item[(iv)]$\STR$ is charge and total four-momentum conserving since its vertices are such. 
\item[(v)]$\STR$ is Lorentz-invariant, since the propagators (\ref{phpro}) and (\ref{elpro}) commute with Lorentz transformations and the vertices are Lorentz-invariant. 
\item[(vi)]In the asymptote $\Lambda \to \infty$, the propagators (\ref{phpro}) and (\ref{elpro}) behave the same way as required by Bogoliubov and Shirkov\cite{Bogol} of the Pauli-Villars--regularized spin 1 and spin $1\over 2$ propagators when their auxiliary masses tend to infinity in renormalization. 
\end{itemize}
So, $\STR$ is a result of a pp-theory as defined in Section~I. This theory can be regarded as a physically plausible regularization of QED based on a physically motivated modification $\LagT$ of QED Lagrangian $\LagQ$.

In view of (vi), we expect that we can compute by renormalization the renormalized n-point Green functions of QED with massless photons from the n-point Green functions of $\LagT$ by choosing an appropriate dependence of $e$, $s_1$, $s\DD$, $q_1$, and $q\DD$ on $\Lambda$, and then limiting $\Lambda \to \infty$ and the renormalized photon mass to zero. However, it is an open question whether the functions $f(p^2)$ and $t(p^2)$ must have also some additional properties for the photon third polarization vector to decouple in the limit of negligible photon mass\cite{Veltm}.

\section{Comments}

Generalizing the 't Hooft and Veltman method of unitary regulators we have shown, for the first time as far as we know, that one can regularize QED in accordance with the basic tenets of theoretical physics by suitably modifying the free part of QED Lagrangian. As we mentioned in Sec.III, the physical motivation for such modification has been the Feynman surmise about X-ons, and the Boltzmann improvement on fluid dynamics by the transport theory based on his equation. 

Within the framework of quantum field theory as defined by 't Hooft and Veltman\cite{Hooft}, the Lagrangian $\LagT$ is related to the physical world solely through the perturbatively defined scattering matrix $\STR$. We see no physical properties of $\STR$ that require the spectral function (\ref{KLrho}) and the Hamiltonians corresponding to free Lagrangians $\Lag_1$ and $\Lag\DD$ (which are not free-particle Lagrangians) to be positive as they turn out to be within the framework of canonical formalism\cite{Cao}.

The need for a regularization of QED that would result in a physically plausible model was felt very strongly by the founders of QED, Dirac and Heisenberg, already some sixty-five years ago\cite{Cao}. But neither they nor their contemporaries succeded in getting rid of the ultraviolet divergencies by a physicaly motivated modification of the QED Lagrangian. In the late 1940s, however, Tomonaga, Schwinger and Feynman ``solved'' the problem of QED ultraviolet divergencies through renormalization---a solution which does not require regularization to be physically plausible, and removes all parameters characteristic of regularization. As they obtained spectacularly succesful formulas for quantum-electrodynamic phenomena, the problem of finding a physically plausible, Lagrangian-based regularization of QED was not so urgent any more. As there had been no progress whatsoever towards a solution of this problem, it mainly came to be considered as practically unsolvable\cite{Cao}; those who hoped otherwise were often considered ``irrational'', as Isham, Salam, and Strathdee\cite{Salam} complained twenty-five years later. Thus nowadays, as far as we know, no quantum-field theorist, excepting the string theorist, pays much attention to this problem, which many of the preceding generations---e.g., Dirac, Heisenberg, Landau, Pauli, and Salam, to mention some---still hoped to be solved somehow someday\cite{Cao}. But the string theorists abandon one of the basic premises of conventional physics, the four-dimensionality of space-time. We have shown, however, that such drastic steps may be avoided when modifying QED Lagrangian to get rid of ultraviolet divergencies. But the question remains which modification of the type considered is the most appropriate for better describing quantum-electrodynamic phenomena at higher energies than the conventional QED, and what is its content.

\section{acknowledgement}

Authors greatly appreciate discussions with M. Polj\v sak and B. Bajc and their suggestions.

\end{document}